\documentclass{article}

\usepackage[toc,page]{appendix}
\usepackage{graphicx}
\usepackage{epstopdf}
\usepackage{amsmath, amsthm, latexsym}
\usepackage{amssymb}
\usepackage{slashed}

\newcommand{\com}{\mathpunct{\raisebox{0.5ex}{,}}}

\usepackage{color}

\usepackage[small, bf]{caption}
\usepackage{subfigure}
\usepackage{hyperref}

\newtheorem{theorem}{Theorem}

\newtheorem{definition}{Definition}
\newtheorem{corollary}{Corollary}
\newtheorem{proposition}{Proposition}
\newtheorem{remark}{Remark}

\newtheorem{example}{Example}

\numberwithin{equation}{section}

\begin{document}

\begin{titlepage}
\begin{center}
\textbf{\Large{Coulomb scattering in the Born approximation and the use of generalized functions}}\\\vspace{6mm}

Peter Collas\\
\vspace{0.5cm}
{\footnotesize Department of Physics and Astronomy,\\
California State University, Northridge, Northridge, CA 91330-8268}\\
{\footnotesize peter.collas@csun.edu}
\vspace{0.5cm}

 {\small (Feb. 23, 2021)} 
 \end{center}

\begin{abstract}
We discuss three ways of obtaining the Born approximation for Coulomb scattering:  The standard way, making use of a convergence factor (``screening''), Oppenheimer's way using cylindrical (instead of spherical) coordinates, and finally Landau and Lifshitz' way.  The last one  although it does require some background from the theory of generalized functions is nevertheless a very instructive and important technique deserving more exposure to physicists.
\end{abstract}

\vspace{1cm}
\noindent {\small KEY WORDS: Potential scattering, Born approximation, Coulomb scattering, Generalized functions, Theory of distributions}\\

\end{titlepage}\,

\section[Introduction]{Introduction}\label{Intro}

Scattering theory is a very important subject since our knowledge about elementary particles and their interactions is based mainly on scattering experiments.  Quantum mechanics texts, e.g., \cite{LL77}, \cite{W15}, \cite{A04}, usually deal with potential scattering, the Born series expansion for the scattering amplitude and work out examples using its lowest term of the series, namely, the Born approximation.  In the present paper we shall be concerned with the calculation of the Born approximation for Coulomb scattering, however we would like to refer the reader to reference \cite{K91} for some applications of the Born approximation in strong interactions.\\

Rutherford in his seminal 1911 paper on atomic structure \cite{R11},  derived the expression for the differential cross section of Coulomb scattering, using only classical mechanics (a derivation may found in ref. \cite{LL76}).  In fact Rutherford introduced the concept of what we now call \textit{the differential cross section} for potential scattering.  As Weinberg points out \cite{W15}, Rutherford was lucky, in that the Coulomb case is the only one for which the classical mechanics result agrees perfectly with the quantum mechanics result!\\

It was not until 1926 that Born \cite{B26I}, \cite{B26II}, using quantum mechanics, derived the series expansion for the scattering amplitude  whose lowest term is the Born approximation, $f_{B}$. Born also noted the fact that each term in this series diverges for the case of the Coulomb potential.  Four months later, Wentzel \cite{W26}, evaluated $f_{B}$, for Coulomb scattering, in spherical coordinates, by making use of a convergence factor.  To our knowledge, \textit{all quantum mechanics texts} perform the integration involved in calculating $f_{B}$, using Wentzel's limiting procedure.  However, in 1927 Oppenheimer \cite{O27}, derived the Born approximation for Coulomb scattering without the use of Wentzel's limiting procedure by performing the integrations in cylindrical coordinates.  Unfortunately, due to his choice of the orientation of the momentum transfer vector $\vec{q}$, his calculation is too complicated and this is probably the reason it has disappeared from the literature.  In this paper we greatly simplify and clarify Oppenheimer's calculation in the hope that this version will be a useful alternative to lecturers on this subject.  Finally we present an evaluation found in a lesser known Landau and Lifshitz text \cite{LL74}.  Although we believe this derivation to be the best, it involves a number of mathematically illegal steps.  We clarify the derivation and supply the necessary generalized function theory machinery thus making all steps rigorous.\\.  

In Sec. \ref{Born I} we review the standard calculation of the Born approximation for Coulomb scattering.  In Sec. \ref{Born II} we give the simplified version of Oppenheimer's approach.  In Section \ref{Born III} we go over the Landau and Lifshitz derivation.  Section \ref{GenFunc} is devoted to an elementary presentation of all the necessary tools from the theory of generalized functions.  Finally in Sec. \ref{Sum}  we sum up our results.  Appendix \ref{UnDim} deals with our choice of units and dimensions, Appendix \ref{Scatt} is a review of potential scattering, and Appendix \ref{Oppen} we give some details of Oppenheimer's original calculation.

\section[The Born approximation I]{The Born approximation I}\label{Born I}

We first outline the standard calculation of the scattering amplitude in the Born approximation.  It is convenient to adopt Planck natural units so that, $c=\hbar=1$, see Appendix \ref{UnDim} for more details.  The momentum transfer vector $\vec{q}$, is given by (see Appendix \ref{Scatt}, Fig. \ref{Fig1})
\begin{equation}
\label{q1}
\vec{q}=\vec{p}_{1}^{\;\prime}-\vec{p}_{1},
\end{equation}

\noindent where $\vec{p}_{1}$ and $\vec{p}_{1}^{\;\prime}$ are, respectively the initial and final momenta of one of the particles in the center of mass frame.  For elastic scattering, $\vert\vec{p}_{1}\vert=\vert\vec{p}_{1}^{\;\prime}\vert= p$, and we have that,
\begin{equation}
\label{q2}
q=2p\sin{\left(\frac{\theta}{2}\right)}\,\com
\end{equation}

\noindent where $\theta\in\left[0,\pi\right]$ and $q\in\left[0,2p\right]$.\\

For a spherically symmetric potential energy, $V(r)$, the scattering amplitude, in the Born approximation, is given by Eq. \eqref{C13} in Appendix \ref{Born}, see e.g., \cite{LL77},
\begin{equation}
\label{fB1.0}
f_{B}=-\frac{m}{2\pi}\int V(r)e^{-i\vec{q}\cdot\vec{r}}d^{3}r,
\end{equation}
 
\noindent where, $m=m_{1}m_{2}/(m_{1}+m_{2})$, is the reduced mass, so that for electron - proton scattering, $m\approx m_{e}$.  When using spherical coordinates, $(r,\vartheta, \phi)$, to evaluate Eq. \eqref{fB1.0}, it is important to bear in mind that the angle $\vartheta$, is not related to the scattering angle $\theta$ in Eq. \eqref{q2}.  In performing the integrations it is most convenient to choose the polar axis in the direction of the vector $\vec{q}$, then the exponent $-i\vec{q}\cdot\vec{r}=-iqr\cos{\vartheta}$.  Integrating over $\vartheta$ and $\phi$, we obtain
\begin{equation}
\label{fB1}
f_{B}=-\frac{2m}{q}\int_{0}^{\infty}V(r)\sin{(qr)}\,r\,dr.
\end{equation}

\noindent  In the Gaussian system of units (see Appendix \ref{UnDim}) the Coulomb potential energy (or Coulomb potential for short) is given by
\begin{equation}
\label{fB1.1}
V(r)=\frac{e_{1}e_{2}}{r}\,\com
\end{equation}

\noindent where $e_{i}=N_{i}(\pm e),\;i=1,2$, $e> 0$ and $N_{i}=$ positive integer.  To keep our notation simple, we will let $N_{i}=1$, thus we write
\begin{equation}
\label{fB1.2}
V(r)=\frac{\pm e^{2}}{r}\,\cdot
\end{equation}

\noindent The signs, $(+,-)$, correspond to repulsion and attraction respectively.\\

The convergence of the integral in Eq. \eqref{fB1}, in spherical coordinates, requires that \cite{AR65},
\begin{equation}
\label{fB1.3}
\int_{0}^{\infty}r^{2}\vert V(r)\vert\,dr<\infty.
\end{equation}

\noindent thus it is evident that $V(r)$ does not vanish sufficiently fast at large $r$ to allow for a straightforward evaluation of the above integral.\\

The universally adopted procedure in the Coulomb case is to define $f_{B}$ using the expression below
\begin{equation}
\label{fB2}
f_{B}:=\lim_{\lambda \,\to \,0}\left(-\frac{2m}{q}\int_{0}^{\infty}V(\lambda,r)\sin{(qr)}\,r\,dr\right),
\end{equation}

\noindent where $V(\lambda,r)$ is often referred to as the \textit{screened} Coulomb potential,
\begin{equation}
\label{fB3}
V(\lambda,r)=\pm e^{2}\left(\frac{e^{-\lambda r}}{r}\right)\,\cdot
\end{equation}

\noindent The parameter $\lambda$, in our units has dimensions of inverse length (or mass), and so it is related to a cutoff length $r_{c}$, by $\lambda=1/r_{c}$.  The exponential factor in Eq. \eqref{fB3} is a qualitative description for the screening of the nuclear charge of an atom by the electrons of the surrounding shells (regarded as a continuous charge density).  A potential of the form \eqref{fB3} is also known as the Yukawa potential because it was proposed by Yukawa \cite{Y35} to describe the strong nucleon-nucleon interaction mediated by the exchange of a spinless boson of mass $\lambda$, (with $g^{2}$ instead of $e^{2}$).\\

Wentzel \cite{W26}, who did the original calculation of the Born approximation for the Coulomb potential,  wrote:  ``So you cannot get by with the pure Coulomb field (our Eq. \eqref{fB1.1}), but at least qualitatively you have to take into account \textit{the screening by the outer electron shells}'' (our italics). Under the circumstances the cutoff $r_{c}$ would be of the order of an atomic radius.  However he added that  ``...the special choice of the exponential function in (our Eq. \eqref{fB3}) is completely irrelevant for the end result; it only offers the advantage of ensuring the convergence of the method in the simplest possible way.''  A lot of research on screening, both theoretical \cite{ST76}, \cite{GT80} and experimental \cite{ABLM80}, \cite{BHN90} has been carried out.\\

Nevertheless it is important to recall that the Born approximation for Coulomb scattering is also valid for electron-electron scattering \textit{where there is no screening in the above sense}.  Here one would have to introduce the idea of vacuum polarization of the virtual pairs and charge renormalization \cite{IZ80}, \cite{V94}, \cite{LB14}, and make an intuitive argument along the lines that effectively, the target charge varies, and
\begin{equation}
\label{fB3.1}
e\rightarrow e\left(e^{-\lambda r}\right).
\end{equation}

\noindent Oppenheimer's approach avoids these problems by obtaining $f_{B}$, without the need of a convergence factor.\\

It is worth pointing out that, higher order terms in the Born series with the potential of Eq. \eqref{fB3}, diverge in the limit $\lambda\rightarrow 0$ \cite{N82}, \cite{D51}, and require a special procedure in order to cancel the divergences \cite{N55}.\\

 We should add that Mott \cite{M28}, and independently Gordon \cite{G28}, calculated the \textit{exact} Coulomb potential scattering amplitude, by solving the Schr\"{o}dinger equation.  Although their approaches are very different, neither of them needed any convergence factor.  Both calculations are rather demanding and Gordon's method, making use of parabolic coordinates, is the one usually worked out but only in more advanced quantum mechanics texts \cite{LL77}, \cite{N82}.  Gordon's solution is another example where a judicious choice of the coordinate system was helpful.\\

In the usual approach, Eq, \eqref{fB2} is the Laplace transform of $\sin{(qr)}$.  The integration is standard and is done by performing two integrations by parts, thus
\begin{align}
f_{B}&=\lim_{\lambda \,\to \,0}\left(\mp\,\frac{2me^{2}}{q}\int_{0}^{\infty}e^{-\lambda r}\sin{(qr)}\,dr\right)\,\com\label{fB4}\\\nonumber\\
&=\lim_{\lambda \,\to \,0}\left(\mp\,\frac{2me^{2}}{q^{2}+\lambda^{2}}\right),\label{fB5}\\\nonumber\\
&=\mp\,\frac{2me^{2}}{q^{2}}\,\com\label{fB6}\\\nonumber\\
&=\mp\,\frac{me^{2}}{2p^{2}\sin^{2}{\left(\frac{\theta}{2}\right)}}\,\com\label{fB7}
\end{align}

\noindent where we have used Eq. \eqref{q2}.

\section[The Born approximation II]{The Born approximation II}\label{Born II}

In this section, following Oppenheimer's idea, we begin with Eq. \eqref{fB1.0} and use cylindrical coordinates $(z, \varrho, \varphi)$.  In contrast with his choice of the orientation of the momentum transfer $\vec{q}$, (see Appendix \ref{Oppen}, Eq. \eqref{O11}), we choose the direction of the vector $\vec{q}$ to be along the positive $z$-axis, thus the exponent in the integrand of Eq. \eqref{fB1.0} is $-i\vec{q}\cdot\vec{r}=-iqz$, which simplifies considerably Oppenheimer's original calculation.  The Coulomb potential $V$, Eq. \eqref{fB1.2}, is now given by
\begin{equation}
\label{f2B0}
V(\varrho,z)=\pm \,\frac{e^{2}}{\sqrt{\varrho^{2}+z^{2}}}\,\cdot
\end{equation}

\noindent The resulting integrals in this case are referred to as \textit{iterated integrals}, i.e., the integrations over $z$ and $\rho$ \textit{have to be done in the order below}, (Oppenheimer does not mention this).  The condition under which one is allowed to interchange the order of integration is given by Fubini's theorem \cite{R87}, and the integrals in Eq. \eqref{f2B1}, fail to satisfy this condition because the integration of the absolute value of the integrand over $z$ does not converge.  Therefore we have that
\begin{align}
f_{B}&:=\mp\frac{me^{2}}{2\pi}\int_{0}^{\infty}\varrho\,d\varrho \int_{-\infty}^{\infty}\frac{e^{-iqz}}{\sqrt{\varrho^{2}+z^{2}}}\,dz \int_{0}^{2\pi}\,d\varphi\,,\label{f2B1}\\\nonumber\\
&=\mp\,me^{2}\int_{0}^{\infty}\varrho\,d\varrho \int_{-\infty}^{\infty}\frac{e^{-iqz}}{\sqrt{\varrho^{2}+z^{2}}}\,dz\,,\label{f2B2}\\\nonumber\\
&=\mp\,2me^{2}\int_{0}^{\infty}\varrho\,K_{0}(q\varrho)\,d\varrho\,, \label{f2B3}\\\nonumber\\
&=\mp\,\frac{2me^{2}}{q^{2}}\,\com\label{f2B4}
\end{align}

\noindent which is again Eq. \eqref{fB6}.  $K_{0}$ is a modified Bessel function of the second kind.  The integrals involved in Eq. \eqref{f2B2} and \eqref{f2B3} may found, for example, in ref. \cite{MOS66}, or may obtained immediately using \textit{Mathematica} \cite{WM20}, or any other symbolic computation software.  In connection with the integral in Eq. \eqref{f2B3}, we remark that although, $\lim_{\varrho \to 0}{\left[K_{0}(q\varrho)\right]}\rightarrow\infty$, it turns out that, $\lim_{\varrho \to 0}{\left[\varrho\,K_{0}(q\varrho)\right]}=0$ so everything is well-behaved.

\section[The Born approximation III]{The Born approximation III}\label{Born III}

In this section we present our final calculation of the Born approximation for the Coulomb case which was given as far as we know only in a less known text by Landau and Lifshitz \cite{LL74}, (not to be confused with \cite{LL77}).  This derivation takes advantage of the fact that the Born approximation of the scattering amplitude, $f_{B}$, Eq. \eqref{fB1.0}, is, apart from an overall constant, the Fourier transform of the potential $V(\vec{r}\,)$.  In our derivation below, we initially assume that the function $g(\vec{r}\,)$ is sufficiently well-behaved so that the various steps in the calculation are permissible.  An example of such a function is $g(\vec{r}\,)=\exp{\left(-r^{2}/2\right)}$.  Then, in Section \ref{GenFunc}, we present a sufficient amount of the theory of generalized functions to enable us to make all the steps mathematically rigorous.\\

The Fourier transform $\widehat{g}(\vec{q}\,)$ of the function $g(\vec{r}\,)$ is given by
\begin{equation}
\label{LL1}
\widehat{g}(\vec{q}\,)=\int g(\vec{r}\,)\,e^{-i\vec{q}\cdot\vec{r}}\,d^{3}r\,.
\end{equation}

\noindent where, in Cartesian coordinates $(x,y,z)$, we have, $d^{3}r=dxdydz$, $d^{3}q=dq_{x}dq_{y}dq_{z}$, and all limits of integration are $-\infty$ to $+\infty$.\\

We recall the integration by parts formula involving two functions $f$ and $\varphi$,
\begin{equation}
\label{LL2}
\int\varphi\,df=f\varphi-\!\int fd\varphi,
\end{equation}

\noindent and remark that if the functions $f$ and $\varphi$ are such that the product $f\varphi$ vanishes at $\pm\infty$ in $R^{3}$, then Eq. \eqref{LL2} reduces to
\begin{equation}
\label{LL3}
\int\varphi\,df=-\!\int fd\varphi.
\end{equation}

\noindent With this in mind, and integrating by parts twice, one can show that the Fourier transform $\widehat{\triangle g}$ of $\triangle g=\left(\partial^{2}_{x}+\partial^{2}_{y}+\partial^{2}_{z}\right)g$, is 
\begin{equation}
\label{LL4}
\int (\triangle g)\,e^{-i\vec{q}\cdot\vec{r}}\,d^{3}r=\int\! g\triangle\!\left(e^{-i\vec{q}\cdot\vec{r}}\right)d^{3}r=-q^{2}\;\widehat{g}(\vec{q}\,),
\end{equation}

\noindent where $q^{2}=q_{x}^{2}+q_{y}^{2}+q_{z}^{2}$.  Then from Eqs. \eqref{LL1} and \eqref{LL4}, we have that
\begin{equation}
\label{LL5}
\widehat{g}(\vec{q}\,)=\int g(\vec{r}\,)\,e^{-i\vec{q}\cdot\vec{r}}\,d^{3}r=-\frac{1}{q^{2}}\int (\triangle g)\,e^{-i\vec{q}\cdot\vec{r}}\,d^{3}r.
\end{equation}

In our case, using Eq. \eqref{fB1.0}, we see that 
\begin{align}
f_{B}&=-\frac{m}{2\pi}\int\frac{\left(\pm e^{2}\right)}{r}\,e^{-i\vec{q}\cdot\vec{r}}d^{3}r,\label{LL6}\\\nonumber\\
&=\mp\left(\frac{me^{2}}{2\pi}\right)\int\frac{e^{-i\vec{q}\cdot\vec{r}}}{r}\,d^{3}r:=\mp\left(\frac{me^{2}}{2\pi}\right)\widehat{g}(\vec{q}\,),\label{LL7}
\end{align}

\noindent and so
\begin{equation}
\label{LL8}
g(\vec{r}\,)=\frac{1}{\vert\vec{r}\,\vert}\com
\end{equation}

\noindent where $\vert\vec{r}\,\vert=r=\sqrt{x^{2}+y^{2}+z^{2}}$.  At this point we use the equation,
\begin{equation}
\label{LL9}
\triangle\frac{1}{\vert\vec{r}\,\vert}=-4\pi\delta (\vec{r}\,).
\end{equation}

\noindent Equation \eqref{LL9} shall be derived in the next section using the appropriate mathematical tools.  Substituting Eq. \eqref{LL9} in Eq. \eqref{LL5}, we find that
\begin{equation}
\label{LL10}
\widehat{g}(\vec{q}\,)=\frac{4\pi}{q^{2}}\com
\end{equation}

\noindent which when substituted in Eq. \eqref{LL7} gives us the desired scattering amplitude 
\begin{equation}
\label{LL11}
f_{B}=\mp\,\frac{2me^{2}}{q^{2}}\,\cdot
\end{equation}

It is clear that in the integration by parts, which led to Eqs. \eqref{LL4}, we neglected a divergent term, furthermore the resulting integrals do not exist!  For the $g(\vec{r}\,)$ of Eq. \eqref{LL8}, the preceding calculations can only be made rigorous in the context of the theory of generalized functions.  In Sec. \ref{GenFunc} we present a concise introduction to the necessary formalism, we derive Eq. \eqref{LL9} and prove Eq. \eqref{LL4}.

\section[Generalized functions]{Generalized functions}\label{GenFunc}

In this Section we present the minimum amount of the theory of generalized functions required in order to put the calculations in Sec. \ref{Born III} on a mathematically sound basis.  In addition we believe that this presentation will make it easier for readers to pursue the subject in any of the excellent textbooks \cite{J82}, \cite{Z87}, \cite{RY90}, \cite{V02}, \cite{K04}.  For a review of generalized functions in connection with applications to electromagnetism we recommend \cite{SW89}.\\

We shall be concerned with applications involving Fourier transforms and for that reason we shall require a set of ``test'' functions $\varphi(x)$ (initally in $R^{1}$) which are called \textit{good or rapidly decreasing test functions}.

\begin{definition}\label{good} A function $\varphi$ is said to be good if $\varphi\in C^{\infty}$ and if
\begin{equation}
\label{D1}
\lim_{|x| \to \infty}\left\vert x^{m}\frac{d^{k}}{dx^{k}}\varphi(x)\right\vert=0,
\end{equation}

\noindent for every integer $m\geq 0$ and every integer $k\geq 0$.  It is evident that if $\varphi$ is good so is $d\varphi/dx$.
\end{definition}

\noindent In the customary notation we write, $\varphi\in S$, where $S$ is the space of good test functions.\\

Now we let $f$ be a functional.  The functional $f$ assigns a number to any good test function $\varphi$ denoted $\langle f,\varphi\rangle$.  If $f$ is a locally integrable function, then we may write
\begin{equation}
\label{D2}
\langle f,\varphi\rangle=\int_{-\infty}^{\infty}f(x)\varphi(x)dx.
\end{equation}

\noindent However if $f$ is not a locally integrable function, then the right-hand side of Eq. \eqref{D2} does not make sense.  In order for $\langle f,\varphi\rangle$ to be a finite number for $\varphi\in S$, it is sufficient, \textit{but not necessary}, to restrict $f(x)$ to the space of functions of slow growth.  This is an important set of functions that will enable us to deal with cases where the right-hand side of Eq. \eqref{D2} does not make sense, and give meaning to  $\langle f,\varphi\rangle$ by introducing the concept of  \textit{generalized functions}, (see Example \ref{delta} below).  The terms \textit{generalized function} and \textit{distribution} will be used interchangeably in what follows.

\begin{definition}\label{slow} A function $f(x)$ is said to be a function of slow growth if, for some  \textup{(}finite\textup{)} integer $N\geq 0$,
\begin{equation}
\label{D3}
\int_{-\infty}^{\infty}\frac{\vert f(x)\vert}{\left(1+x^{2}\right)^{N}}\,dx<\infty.
\end{equation}
\end{definition}

\noindent We say that $f\in K$.  The generalization of Definition \ref{slow} to $n$ dimensions is straightforward, $x^{2}\rightarrow r^{2}=x_{1}^{2}+\dots+x_{n}^{2}$, etc.  For example (for $n=3$) the function $f=(1/r)\in K$ (use spherical coordinates).\\

\noindent Note that good functions decrease faster than any power of $\vert x\vert$ as $x\rightarrow\pm\infty$, e.g., $\exp{(-x^{2})}$, (but note that $\exp{(-\vert x\vert)}$ is not a good function). Functions $f$ of slow growth grow at infinity like polynomials, e.g., $\exp{(ix)}$. 

\begin{remark}\label{fphi}  An important and rather obvious consequence is a theorem that states that the product of a function $f$ of slow growth and a rapidly decreasing function $\varphi$, is a rapidly decreasing function $f\varphi$.
\end{remark}

\begin{definition}\label{temp} The piecewise continuous function of slow growth $f(x)$, defines the tempered distribution\textup{ (}or distribution of slow growth\textup{)}
\begin{equation}
\label{D4}
\langle f,\varphi\rangle=\int_{-\infty}^{\infty}f(x)\varphi(x)dx,
\end{equation}
for all good functions $\varphi$.
\end{definition}

\noindent The set of all tempered distributions is denoted by $S^{\prime}$.\\

We can now differentiate $f(x)$ using integration by parts,
\begin{equation}
\label{D5}
\int_{-\infty}^{\infty}\frac{df(x)}{dx}\varphi(x)dx=\left[f(x)\varphi{(x)}\right]_{-\infty}^{\infty}-\int_{-\infty}^{\infty}f(x)\frac{d\varphi(x)}{dx}dx.
\end{equation}

\noindent It follows from Remark \ref{fphi} that $\left[f(x)\varphi{(x)}\right]_{-\infty}^{\infty}=0$, therefore we simply have that
\begin{align}
\langle f^{\prime},\varphi\rangle&=\!-\langle f,\varphi^{\prime}\rangle,\label{D6}\\\nonumber\\
\langle f^{\prime\prime},\varphi\rangle&=\!-\langle f^{\prime},\varphi^{\prime}\rangle=\langle f,\varphi^{\prime\prime}\rangle,\label{D7}
\end{align}

\noindent and so forth.  Note that Eqs. \eqref{D6} and \eqref{D7} hold even if $f(x)$ is not differentiable (since $\varphi$ is).

\begin{remark}\label{diff} It is easy to see that the generalized derivative, Eq. \eqref{D6}, of a generalized function, $f$, is also a generalized function, $f^{\prime}$.
\end{remark}

\begin{example}\label{delta} As our first example we show how to obtain the Dirac $\delta$ function as the derivative of the unit step function
\begin{equation}
\label{D8}
\theta (x)=
\begin{cases}
1, &x\geq 0\\
0, &x<0\,.\\ 
\end{cases}
\end{equation}
\end{example}

\noindent Although the derivative, $\theta^{\prime}(0)$ does not exist in the usual sense.  $\theta(x)\in K$ and $\varphi\in S$, therefore from Eq. \eqref{D6}, we have
\begin{align}
\langle \theta^{\prime},\varphi\rangle&=-\!\int_{-\infty}^{\infty}\theta(x)\frac{d\varphi}{dx}\,dx,\label{D9}\\\nonumber\\
&=-\!\int_{0}^{\infty}\frac{d\varphi}{dx}\,dx,\label{D10}\\\nonumber\\
&=-\varphi(x)\vert_{0}^{\infty}=\varphi(0):=\langle \delta,\varphi\rangle\in S^{\prime}. \label{D11}
\end{align}

\noindent So $\theta^{\prime}(x)$ maps every test function $\varphi(x)$ to its value at the origin and enables us to define the generalized function $\delta(x)$.\\

We now turn our attention to the Fourier transform.  We recall that the Fourier transform $\widehat{f}(x)$ (or $\left[f(x)\right]^{\wedge}$), of a well-behaved function $f(x)$, is
\begin{equation}
\label{D12}
\widehat{f}(q)=\int_{-\infty}^{\infty} f(x)\,e^{-iqx}\,dx\,.
\end{equation}

\noindent We need to make use of \textit{Parseval's equation}, which for well-behaved functions  $f$ and $g$, is easy to prove using Fubini's theorem \cite{R87},
\begin{align}
\int_{-\infty}^{\infty}\widehat{f}(q)g(q)dq&=\int_{-\infty}^{\infty}\left[\int_{-\infty}^{\infty}f(x)e^{-iqx}dx\right]g(q)dq,\label{D13}\\\nonumber\\
&=\int_{-\infty}^{\infty}f(x)\left[\int_{-\infty}^{\infty}g(q)e^{-iqx}dq\right]dx,\label{D14}\\\nonumber\\
&=\int_{-\infty}^{\infty}f(x)\widehat{g}(x)dx.\label{D15}
\end{align}

\begin{definition}\label{ft} Let $f(x)$ be a piecewise continuous function of slow growth, then we use Parseval's equation to define the Fourier transform $\widehat{f}$, of the generalized function $f$, to be 
\begin{equation}
\label{D16}
\langle\widehat{f},\varphi\rangle:=\langle f,\widehat{\varphi}\rangle,
\end{equation}
for all good functions $\varphi$.
\end{definition}

\noindent In order for Definition \ref{ft} to be meaningful we need the theorem below which we state without proof.

\begin{theorem}\label{goodft}
If $\varphi(x)$ is a good function so is its Fourier transform $\widehat{\varphi}(q)$.
\end{theorem}

\noindent This is the reason why the test functions $\varphi$ had to be good.  From Theorem \ref{goodft} and Definition \ref{temp} it follows that if $f$ is a function of slow growth, then its Fourier transform $\widehat{f}$ is a tempered distribution (generalized function of slow growth).\\

\noindent Since $\varphi$ is a good function, it is easy to deduce by direct calculation that for $k$ a non-negative integer
\begin{align}
\partial^{k}\left(\widehat{\varphi}\right)&=\left[(-ix)^{k}\varphi\right]^{\wedge},\label{D17}\\\nonumber\\
\left[\partial^{k}\varphi\right)]^{\wedge}&=(iq)^{k}\widehat{\varphi}.\label{D18}
\end{align}

\begin{example}\label{delta2} We show that the Fourier transform of the $\delta$ function is 1.
\end{example}
\begin{equation}
\label{D19}
\langle\widehat{\delta},\varphi\rangle=\langle\delta,\widehat{\varphi}\rangle=\widehat{\varphi}(0)=\int_{-\infty}^{\infty}\varphi(x)dx=\langle 1,\varphi\rangle,
\end{equation}

\noindent where we have used Eq. \eqref{D11}.

\begin{remark}\label{gphi} The above example is simple but it is important to note that every step in Eqs \eqref{D19} is independent of the particular $\varphi$.
\end{remark}

\noindent Using Eqs. \eqref{D15}, \eqref{D6} and \eqref{D7} we obtain two very handy results.

\begin{corollary}\label{dft}
\begin{equation}
\label{D20}
\langle\widehat{(f^{\prime})},\varphi\rangle=\langle f^{\prime},\widehat{\varphi}\rangle=-\langle f,\left(\widehat{\varphi}\right)^{\prime}\rangle.
\end{equation}
 \end{corollary}
 
\begin{corollary}\label{2dft}
\begin{equation}
\label{D21}
\langle\widehat{(f^{\prime\prime})},\varphi\rangle=\langle f^{\prime\prime},\widehat{\varphi}\rangle=-\langle f',\left(\widehat{\varphi}\right)^{\prime}\rangle=\langle f,\left(\widehat{\varphi}\right)^{\prime\prime}\rangle.
\end{equation}
 \end{corollary}
 
\noindent We re-write Eq. \eqref{D21} for clarity
\begin{equation}
\label{D22}
\int_{-\infty}^{\infty}\left[\frac{d^{2}}{dx^{2}}f(x)\right]^{\wedge}(q)\varphi(q)dq=\!\int_{-\infty}^{\infty}f(x)\left(\frac{d^{2}}{dx^{2}}\,\widehat{\varphi}(x)\right)dx.
\end{equation}

\begin{remark}\label{Rn} All of the preceding results in this Section can be extended \textup{(}with appropriate minor changes\textup{)} to several dimensions, that is, $x\in R^{n}$.  Thus, for example, for $x\in R^{n}$ Eqs. \eqref{D7} and \eqref{D21} become
\begin{align}
\langle\triangle f,\varphi\rangle&=\langle f,\triangle\varphi\rangle,\label{D23}\\\nonumber\\
\langle\widehat{(\triangle f)},\varphi\rangle&=\langle f,\triangle\left(\widehat{\varphi}\right)\rangle,\label{D24}
\end{align}

\noindent where
\begin{equation}
\label{D25}
\triangle=\left(\partial^{2}_{1}+\dots+(\partial^{2}_{n}\right).
\end{equation}
\end{remark}

The first step in the derivation of Eq. \eqref{LL9} is to prove the following theorem.

\begin{theorem}\label{Lapf}
\begin{equation}
\label{D26}
\int_{R^{3}}(\triangle f)\,\varphi\,d^{3}r=\int_{R^{3}}\frac{f}{r^{2}}\frac{\partial}{\partial r}\left(r^{2}\,\frac{\partial\varphi}{\partial r}\right)d^{3}r,
\end{equation}

\noindent where $f=f(r)\in K$, \textup{(}a function of slow growth\textup{)}, $\varphi=\varphi\left(r,\theta_{1},\theta_{2}\right)\in S$, \textup{(}a good function\textup{)}, and $d^{3}r:=r^{2}drd\Omega=r^{2}dr\sin{\theta_{1}}d\theta_{1}d\theta_{2}$.
\end{theorem}

\noindent Proof:

\begin{remark}\label{gprod} In ref. \textup{\cite{J82}} it shown that in $R^{n}$ it is sufficient to work with good functions which are the product of $n$ good functions $\varphi_{k}$, each of which is a function of a single variable.
\end{remark}

\noindent Thus in $R^{3}$  we may write, without loss of generality,
\begin{align}
\varphi&=\varphi\left(r,\theta_{1},\theta_{2}\right)=\varphi(x,y,z)=\varphi_{1}(x)\varphi_{2}(y)\varphi_{3}(z),\label{D27}\\\nonumber\\
&=\varphi_{1}(r\sin{\theta_{1}}\cos{\theta_{2}})\varphi_{2}(r\sin{\theta_{1}}\sin{\theta_{2}})\varphi_{3}(r\cos{\theta_{1}}).\label{D28}
\end{align}

\noindent We shall make use of the fact that $\varphi$ is periodic in $\theta_{2}$, and
\begin{equation}
\label{D29}
\lim_{r \to 0}\varphi\left(r,\theta_{1},\theta_{2}\right):=\varphi(0).
\end{equation}

\noindent From Eq. \eqref{D23} we have that
\begin{equation}
\label{D30}
\int_{R^{3}}(\triangle f)\varphi\,d^{3}r=\int_{R^{3}}f\triangle\varphi\,d^{3}r,
\end{equation}

\noindent and, in spherical coordinates,
\begin{equation}
\label{D31}
\triangle\varphi=A+B+C\,,
\end{equation}

\noindent where
\begin{align}
A&=\frac{1}{r^{2}}\frac{\partial}{\partial r}\left(r^{2}\,\frac{\partial\varphi}{\partial r}\right)\com\label{D32}\\\nonumber\\
B&=\frac{1}{r^{2}\sin{\theta_{1}}}\frac{\partial}{\partial\theta_{1}}\left(\sin{\theta_{1}}\frac{\partial\varphi}{\partial\theta_{1}}\right)\com\label{D33}\\\nonumber\\
C&=\frac{1}{r^{2}\sin^{2}{\theta_{1}}}\frac{\partial^{2}\varphi}{\partial\,\theta^{2}_{2}}\cdot\label{D34}
\end{align}

\noindent Thus we write
\begin{equation}
\label{D35}
\int_{R^{3}}f\triangle\varphi\,d^{3}r=\int_{R^{3}}f(A+B+C)r^{2}drd\Omega,
\end{equation}

\noindent and consider the third term on the right-hand side,
\begin{align}
&\int_{R^{3}}fCr^{2}drd\Omega\label{D36}\\\nonumber\\
&=\int\dots\int_{0}^{2\pi}\frac{\partial^{2}\varphi}{\partial\,\theta^{2}_{2}}\,d\theta_{2}=\int\dots\left[\frac{\partial\varphi}{\partial\theta_{2}}\right]_{0}^{2\pi}=0,\label{D37}
\end{align}

\noindent because $\varphi$ has period $2\pi$ in $\theta_{2}$, (see Eq. \eqref{D28}).  Now we consider the second term on the right-hand side of Eq. \eqref{D35},
\begin{align}
&\int_{R^{3}}fBr^{2}drd\Omega\label{D38}\\\nonumber\\
&=\int\dots\int_{0}^{\pi}\frac{\partial}{\partial\theta_{1}}\left(\sin{\theta_{1}}\frac{\partial\varphi}{\partial\theta_{1}}\right)d\theta_{1},\label{D39}\\\nonumber\\
&=\int\dots\left[\sin{\theta_{1}}\frac{\partial\varphi}{\partial\theta_{1}}\right]_{0}^{\pi}=0\,,\label{D40}
\end{align}

\noindent since $\partial\varphi/\partial\theta_{1}$ is bounded and continuous.  Therefore,
\begin{equation}
\label{D41}
\int_{R^{3}}(\triangle f)\varphi\,d^{3}r=\int_{R^{3}}fA\,d^{3}r,
\end{equation}

\noindent which completes the proof of Eq. \eqref{D26}.\\

We now proceed to the final step required to prove Eq. \eqref{LL9}.  We let
\begin{equation}
\label{D42}
f=\frac{1}{r}\com\;\;\;\;\Rightarrow\;\;\;\;\frac{\partial f}{\partial r}=-\frac{1}{r^{2}}\cdot
\end{equation}

Recall that $f\in K$.  Then integrating by parts Eq. \eqref{D41}, we have
\begin{align}
\int_{R^{3}}(\triangle f)\varphi\,d^{3}r&=\int_{R^{3}}fA\,d^{3}r\label{D43}\\\nonumber\\
&=\int_{R^{3}}f\frac{\partial}{\partial r}\left(r^{2}\frac{\partial\varphi}{\partial r}\right)drd\Omega,\label{D44}
\\\nonumber\\
&=\int d\Omega\left[fr^{2}\frac{\partial\varphi}{\partial r}\right]_{0}^{\infty}-\int_{R^{3}}\frac{\partial f}{\partial r}\,r^{2}\frac{\partial\varphi}{\partial r}d^{3}r.\label{D45}\\\nonumber\\
&=\int_{R^{3}}\frac{\partial\varphi}{\partial r}\,drd\Omega=\int_{0}^{2\pi}\!\int_{0}^{\pi}\left[\varphi\right]_{0}^{\infty}d\Omega\label{D46}\\\nonumber\\
&=-\varphi(0)\int_{0}^{2\pi}\!\int_{0}^{\pi}d\Omega=-4\pi\varphi(0).\label{D47}
\end{align}

\noindent where we have used Eq. \eqref{D29}.  Using the generalization of Eq. \eqref{D11}, we have shown that
\begin{equation}
\label{D48}
\int_{R^{3}}(\triangle f)\varphi\,d^{3}r=-4\pi\int_{R^{3}}\delta(\vec{r}\,)\varphi(\vec{r}\,)d^{3}r=-4\pi\varphi(0).
\end{equation}

\noindent It is in this sense that we may write
\begin{equation}
\label{D49}
\triangle\left(\frac{1}{r}\right)=-4\pi\delta(\vec{r}\,).
\end{equation}

\begin{remark}\label{friedman}  We believe that the following statement about $\delta(x)$ from Friedman's early text \textup{\cite{F56}} captures the essence of generalized functions:  ``We notice that the function $\delta(\vec{r}\,)$ is treated exactly as if it were an ordinary function except that we shall never talk about the ``values'' of $\delta(\vec{r}\,)$.  We talk about the values of integrals involving $\delta(\vec{r}\,)$''.
\end{remark}

We do need to prove one last proposition, namely, Eq. \eqref{LL4}.
\begin{proposition}\label{prop1}  If $f\in K$, then
\begin{equation}
\label{D50}
\left[\triangle f\right]^{\wedge}(\vec{q}\,)=-q^{2}\widehat{f}(\vec{q}\,).
\end{equation}
\end{proposition}

\noindent Proof:\\

\noindent We shall make use of the generalization of Eq. \eqref{D17} to $R^{3}$, namely,
\begin{equation}
\label{D51}
\triangle\widehat{\varphi}=\left[-q^{2}\varphi\right]^{\wedge}.
\end{equation}

\noindent Thus
\begin{align}
\langle\widehat{\triangle f},\varphi\rangle&=\langle\triangle f,\widehat{\varphi}\rangle=\langle f,\triangle\widehat{\varphi}\rangle,\label{D52}\\
&=\langle f,\left[-q^{2}\varphi\right]^{\wedge}\rangle=\langle\widehat{f},-q^{2}\varphi\rangle,\label{D53}\\
&=\langle -q^{2}\widehat{f},\varphi\rangle,
\end{align}

\noindent which is Eq. \eqref{D50} or \eqref{LL4}.

\section[Summary]{Summary}\label{Sum}

We saw that in Wentzel's approach in Sec. \ref{Born I}, Eqs. \eqref{fB2}, \eqref{fB3},  the prescription is:  Perform the integration first and then take the limit $\lambda\rightarrow 0$.  In Oppenheimer's approach Sec. \ref{Born II}, Eq. \eqref{f2B2}, the prescription is:  Do the iterated integrals in the prescribed order.  There is no escaping the fact that the Coulomb potential does not satisfy Eq. \eqref{fB1.3} and consequently evaluating  the Born approximation requires special care.  We hope that our simplification and clarification of Oppenheimer's calculation will be be a useful addition to the usual textbook presentations on this subject.  However we believe that Landau and Lifshitz derivation using the theory of generalized functions is the most satisfactory both from the mathematical and physics point of view since it does not require any extraneous assumptions.

\section[Acknowledgments]{Acknowledgments}\label{Ack}

The author is grateful to David Klein for important comments regarding the material in Section \ref{GenFunc} and Victor Gilinsky for many helpful suggestions and encouragement. 

\begin{appendices}

\section[Units and dimensions]{Units and dimensions}\label{UnDim}

We use Planck natural units, $c=\hbar=1$, and with respect to the electromagnetic  equations, Gaussian units.  Thus the Coulomb potential energy for two simple charges, each $\pm e,\;e>0$, is
\begin{equation}
\label{A1}
V(r)=\frac{\pm e^{2}}{r}\,\com
\end{equation}

\noindent where the signs, $(+,-)$, correspond to repulsion and attraction respectively.  In our units the fine structure constant is dimensionless and
\begin{equation}
\label{A2}
\frac{e^{2}}{\hbar c}\cong\frac{1}{137}\cdot
\end{equation}

\noindent From the Compton wavelength relation below,
\begin{equation}
\label{A3}
\lambda_{C}=\frac{2\pi\hbar}{mc}=\frac{2\pi\hbar c}{mc^{2}}\com
\end{equation}

\noindent it follows that in natural units all quantities have dimensions of length, $L$, or inverse length.  In particular $[E]=[p]=[m]=L^{-1}$.\\

It is also useful to keep in mind that since the differential cross section has dimensions of area,
\begin{equation}
\label{A4}
\left[\frac{d\sigma}{d\Omega}\right]=L^{2},
\end{equation}

\noindent we have that the scattering amplitude has dimensions of length,
\begin{equation}
\label{A5}
\left[f(p,\theta)\right]=L\,.
\end{equation}

\section[Review of scattering theory]{Review of scattering theory}\label{Scatt}

In this Appendix we outline some relevant material leading to the Schr\"{o}dinger equation for elastic scattering by a \textit{central potential} \cite{LL77},  \cite{W15}, \cite{T72}, \cite{OF63}, the definition of the scattering amplitude, $f(p,\theta)$ and the Born approximation, $f_{B}(p,\theta)$.

\subsection[The scattering amplitude]{The scattering amplitude}\label{Psi}

\noindent In the center of mass system we let $\vec{p}_{1},\;\vec{p}_{2}$ be the incoming particle momenta, and $\vec{p}^{\;\prime}_{1},\,\vec{p}^{\;\prime}_{2}$ the outgoing particle momenta.  In elastic scattering the incoming and outgoing particles are the same, moreover their kinetic energies must be equal before and after the collision, thus it follows that
\begin{equation}
\label{B1}
p=\vert\vec{p}_{1}\vert=\vert\vec{p}_{2}\vert=\vert\vec{p}_{1}^{\;\prime}\vert=\vert\vec{p}_{2}^{\;\prime}\vert\,.
\end{equation}

\begin{figure}[!h]
  \begin{center}
    \includegraphics[width=2.8 in]{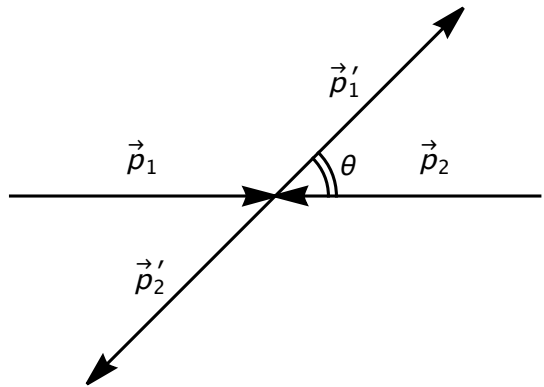}
     \end{center}
   \caption{Ingoing and outgoing momentum vectors for elastic scattering in the center of mass frame.}
  \label{Fig1}
\end{figure}

\noindent We also define the momentum transfer vector $\vec{q}$,
\begin{equation}
\label{B2}
\vec{q}=\vec{p}_{1}^{\;\prime}-\vec{p}_{1},
\end{equation}

\noindent and using Eq. \eqref{B1}, we have that
\begin{equation}
\label{B3}
q^{2}=2p^{2}(1-\cos{\theta}),
\end{equation}

\noindent where $\theta$ is the the center of mass scattering angle, $\theta\in\left[0,\pi\right]$.  In Fig. \ref{Fig1} we show the ingoing and outgoing momenta for elastic scattering in the center of mass frame.\\

\noindent In the center of mass the Schr\"{o}dinger equation for the wavefunction $\psi$ is
\begin{equation}
\label{B4}
\triangle\psi(\vec{r})+2m\left(E-V(r)\right)\psi(\vec{r})=0,
\end{equation}

\noindent where $\triangle$ is the Laplacian in spherical coordinates $(r,\theta,\phi)$.  Since our potential depends only on $r$, the wavefunction $\psi$ will be independent of the azimuthal angle $\phi$.  The energy $E=p^{2}/2m$.  The mass $m$ is the \textit{reduced mass} of the system,
\begin{equation}
\label{B5}
m=\frac{m_{1}m_{2}}{m_{1}+m_{2}}\cdot
\end{equation}

In the time-independent description of scattering, in the asymptotic region, where $V(r)\simeq 0$, the total wavefunction consists of three parts.  Two plane waves, one representing the incident particles and one with the same momentum corresponding to the unscattered particles, these are indistinguishable in the time-independent description and so we write
\begin{equation}
\label{C1}
\psi_{in}=e^{i\vec{p}\cdot\vec{r}}=e^{ipz},
\end{equation}
 \noindent where we have chosen $z$ to be along the incident beam direction (the polar axis).  The third part of the wavefunction is an outgoing spherical wave corresponding to the scattered particles, which we write as,
\begin{equation}
\label{C2}
\psi_{sc}\approx f(p,\theta)\frac{e^{ipr}}{r}\cdot
\end{equation}

\noindent We point out the following useful relations,
\begin{align}
&\vec{p}_{1}^{\;\prime}=p\,\frac{\vec{r}}{r}\com\label{C2.1}\\\nonumber\\
&\vec{p}_{1}^{\;\prime}\cdot \vec{r}=p\,r,\label{C2.2}\\\nonumber\\
&\vec{p}_{1}\cdot\vec{p}_{1}^{\;\prime}=p^{2}\cos{\theta}.\label{C2.3}
\end{align}

\noindent In fact then,
\begin{equation}
\label{C3}
\psi(\vec{r})=e^{ipz}+f(p,\theta)\frac{e^{ipr}}{r}+O\left(\frac{1}{r^{2}}\right).
\end{equation}

\noindent The function $f(p,\theta)$ is \textit{the scattering amplitude} and is in turn related to \textit{the differential scattering   section} by the relation
\begin{equation}
\label{C4}
\frac{d\sigma}{d\Omega}=\vert f(p,\theta)\vert^{2}\,.
\end{equation}

\noindent The differential cross section $d\sigma/d\Omega$ is equal to  the \textit{number of particles \textup{(}of a given kind\textup{)} scattered in the direction $\theta$, per unit solid angle, per unit time, per unit incident flux, per scatterer.}\\

Although we have skipped a lot of important details, we would like to mention that the normalization of the asymptotic form of $\psi$ in Eq. \eqref{C3} was chosen so as to agree with our definition of $d\sigma/d\Omega$.  In summary, we see that in order to obtain $f(p,\theta)$, we have to solve the Schr\"{o}dinger equation for $\psi$, and then find its asymptotic form, Eq. \eqref{C3}.\\

We should add that the case of identical particles requires special care, since $\psi$ has to be made symmetric or antisymmetric depending on whether the total spin of the incoming particles is even or odd respectively.  We refer the interested reader to the best treatment on this subject, namely, section 137 of \cite{LL77}.

\subsection[The Born approximation]{The Born approximation}\label{Born}

Schr\"{o}dinger's equation \eqref{B4} can be converted into the integral equation \cite{OF63}
\begin{equation}
\label{C6}
\psi(\vec{r})=e^{i\vec{p}\cdot\vec{r}}-\frac{m}{2\pi}\int\frac{e^{ip\vert\vec{r}-\vec{r}^{\,\prime}\vert}}{\vert\vec{r}-\vec{r}^{\,\prime}\vert}V(r^{\prime})\psi(\vec{r}^{\;\prime})d^{3}r^{\prime},
\end{equation}

\noindent where $V(r^{\prime})$ is again a spherically symmetric potential.  The integral equation \eqref{C6} is exact and has the advantage that it is constructed so that the ``boundary condition''  of Eq. \eqref{C3} is satisfied automatically by the solution $\psi(\vec{r})$.  It was shown by Born \cite{B26II}, that the above integral equation may be solved by iterating $\psi$.  The resulting series is called the Born series.\\

In the present paper we are concerned with the lowest order term in the Born series, the so-called \textit{Born approximation}.  In order to calculate this term, we have to make some reasonable approximations to Eq. \eqref{C6}.  If we assume that $V(r^{\prime})$ decreases sufficiently rapidly, as $r^{\prime}$ becomes large, then the domain of integration over $r^{\prime}$ is essentially finite and for $r$ very large, we can write
\begin{equation}
\label{C7}
{\vert\vec{r}-\vec{r}^{\,\prime}\vert}=r\left(1+\frac{(r^{\prime})^{2}}{r^{2}}-\frac{2\,\vec{r}\cdot\vec{r}^{\;\prime}}{r^{2}}\right)^{1/2}\approx \,r-\frac{\vec{r}\cdot\vec{r}^{\;\prime}}{r}\cdot
\end{equation}

\noindent Although  we may further approximate, ${\vert\vec{r}-\vec{r}^{\,\prime}\vert}\sim r$ in the denominator of the integrand, the exponential depends sensitively $r^{\prime}$, no matter how large $r$ is.  Nonetheless, using Eq. \eqref{C2.1}, we may write
\begin{equation}
\label{C8}
e^{ip\vert\vec{r}-\vec{r}^{\,\prime}\vert}\approx e^{ip\left(r-\frac{\vec{r}\cdot\vec{r}^{\;\prime}}{r}\right)}=e^{ip\,r-i\vec{p}_{1}^{\;\prime}\cdot\vec{r}^{\;\prime}}.
\end{equation}

\noindent Then, using Eqs. \eqref{C7} and \eqref{C8}, we can simplify Eq. \eqref{C6}
\begin{equation}
\label{C9}
\psi(\vec{r})\approx e^{i\vec{p}\cdot\vec{r}}-\frac{m}{2\pi}\left(\frac{e^{ipr}}{r}\right)\!\int e^{-i\vec{p}_{1}^{\;\prime}\cdot\vec{r}^{\;\prime}}V(r^{\prime})\psi(\vec{r}^{\;\prime})d^{3}r^{\prime},
\end{equation}

\noindent and comparing with Eq. \eqref{C3}, we obtain
\begin{equation}
\label{C10}
f(p,\theta)=-\frac{m}{2\pi}\!\int e^{-i\vec{p}_{1}^{\;\prime}\cdot\vec{r}^{\;\prime}}V(r^{\prime})\psi(\vec{r}^{\;\prime})d^{3}r^{\prime}.
\end{equation}

\noindent It is important to realize that although we did some approximations in obtaining the asymptotic expression of Eq. \eqref{C6}, the expression for the scattering amplitude $f(p,\theta)$ above \textit{is exact}.  The lowest order approximation for $\psi(\vec{r}^{\;\prime})$ is (see Eq. \eqref{C9}),
\begin{equation}
\label{C11}
\psi(\vec{r}^{\;\prime})\sim e^{i\vec{p}_{1}\cdot\vec{r}^{\;\prime}},
\end{equation}

\noindent thus the scattering amplitude Eq. \eqref{C10} becomes
\begin{align}
f_{B}(p,\theta)&=-\frac{m}{2\pi}\!\int e^{-i\left(\vec{p}_{1}^{\;\prime}-\vec{p}_{1}\right)\cdot\vec{r}^{\;\prime}}V(r^{\prime})d^{3}r^{\prime}, \label{C12}\\\nonumber\\
&=-\frac{m}{2\pi}\!\int e^{-i\vec{q}\cdot\vec{r}^{\;\prime}}V(r^{\prime})d^{3}r^{\prime},\label{C13}
\end{align}

\noindent where we have used the momentum transfer, $\vec{q}$, Eq. \eqref{B2}.  At this point we may drop the primes, without any danger of confusion, and obtain our Eq. \eqref{fB1.0}.

\section[Oppenheimer's choice]{Oppenheimer's choice}\label{Oppen}

Before proceeding we remind the reader  that the angle $\vartheta$, in the spherical coordinates below, is not related to the scattering angle $\theta$ in Eq. \eqref{q2}.  We  now consider the momentum transfer vector $\vec{q}$ and the radial vector $\vec{r}$, with Cartesian components,
\begin{align}
\vec{q}&=\left(q^{x},\,q^{y},\,q^{z}\,\right),\label{O1}\\
\vec{r}&=(x,\,y,\,z\,).\label{O2}
\end{align}

\noindent If we choose $\vec{q}$ to be along the polar axis (i.e., the positive $z$-axis) then
\begin{equation}
\label{O3}
\vec{q}=(0,\,0,\,q\,).
\end{equation}

\noindent The Cartesian components of $\vec{r}$ expressed in spherical coordinates,$(r,\vartheta, \phi)$, are
\begin{equation}
\label{O4}
\vec{r}=(r\sin{\vartheta}\cos{\phi},\,r\sin{\vartheta}\sin{\phi},\,r\cos{\vartheta}\,),
\end{equation}

\noindent thus
\begin{equation}
\label{O5}
\vec{q}\cdot\vec{r}=q\,r\cos{\vartheta}.
\end{equation}

Likewise if, instead,we express the Cartesian components of $\vec{r}$ in cylindrical coordinates, $(z, \varrho, \varphi)$, we have
\begin{equation}
\label{O6}
\vec{r}=(\varrho \cos{\varphi},\,\varrho\sin{\varphi},\,z\,),
\end{equation}

\noindent we have
\begin{equation}
\label{O7}
\vec{q}\cdot\vec{r}=q\,z.
\end{equation}
 
 Oppenheimer \cite{O27} chose $\vec{p}_{1}$ to be along the positive $z$-axis, so that
 \begin{equation}
\label{O8}
\vec{p}_{1}=(0,\,0,\,p\,).
\end{equation}

\noindent Then $\vec{p}_{1}^{\,\prime}$, without loss of generality, may be written as (see Fig. \ref{Fig1})
\begin{equation}
\label{O9}
\vec{p}_{1}^{\,\prime}=(p\sin{\theta},\,0,\,p\cos{\theta}\,),
\end{equation}

\noindent hence
\begin{equation}
\label{O10}
\vec{q}=(p\sin{\theta},\,0,\,p\cos{\theta}-p\,).
\end{equation}

\noindent So using Eq. \eqref{O6} we obtain
\begin{equation}
\label{O11}
\vec{q}\cdot\vec{r}=p\varrho\sin{\theta}\cos{\varphi}+p(\cos{\theta}-1)z.
\end{equation}

\noindent As the reader may check by referring to \cite{O27}, the above choice complicates his calculation considerably.

\end{appendices}

\end{document}